\documentclass[conference]{IEEEtran}
\IEEEoverridecommandlockouts
\ifCLASSINFOpdf
\else
\usepackage[dvips]{graphicx}
\fi
\usepackage{url}

\usepackage{cite}
\usepackage{graphicx,amsmath,amssymb,tikz,psfrag}
\usepackage{color}
\usepackage{fancyhdr} 
\usepackage{tikz}
\usetikzlibrary{shapes,arrows,chains}
\usepackage{xcolor}
\usepackage{algorithm}
\usepackage{algpseudocode}
\usepackage{bm}
\usepackage{enumerate}

\begin{document}

\title{Accelerating Edge Intelligence via Integrated Sensing and Communication}

\author{\IEEEauthorblockN{Tong Zhang\textsuperscript{$1$},  Shuai Wang\textsuperscript{$2$}, Guoliang Li\textsuperscript{$1$}, Fan Liu\textsuperscript{$1$}, Guangxu Zhu\textsuperscript{$3$}, and Rui Wang\textsuperscript{$1$}
	}%

	\IEEEauthorblockA{\textsuperscript{$1$}Department of Electrical and Electronic Engineering, Southern University of Science and Technology, Shenzhen, China}
	\IEEEauthorblockA{\textsuperscript{$2$}Shenzhen Institute of Advanced Technology, Chinese Academy of Sciences, Shenzhen, China}
	\IEEEauthorblockA{\textsuperscript{$3$}Shenzhen Research Institute of Big Data, Shenzhen, China}
	Email: \{zhangt7, ligl2020, liuf6, wang.r\}@sustech.edu.cn, s.wang@siat.ac.cn, and gxzhu@sribd.cn 

	\thanks{This work was supported in part by the National Natural Science Foundation of China under Grant 62001203, in part by the Guangdong Basic and Applied Basic Research Project under Grant 2021B1515120067, in part by the Shenzhen Science and Technology Program under Grant RCB20200714114956153. The corresponding author is S. Wang.}
}
\maketitle

\begin{abstract}
Realizing edge intelligence consists of sensing, communication, training, and inference stages. Conventionally, the sensing and communication stages are executed sequentially, which results in excessive amount of dataset generation and uploading time. 
This paper proposes to accelerate edge intelligence via integrated sensing and communication (ISAC). As such, the sensing and communication stages are merged so as to make the best use of the wireless signals for the dual purpose of dataset generation and uploading. However, ISAC also introduces additional interference between sensing and communication functionalities. To address this challenge, this paper proposes a classification error minimization formulation to design the ISAC beamforming and time allocation. The globally optimal solution is derived via the rank-1 guaranteed semidefinite relaxation, and performance analysis is performed to quantify the ISAC gain over that of conventional edge intelligence.
Simulation results are provided to verify the effectiveness of the proposed ISAC-assisted edge intelligence system. Interestingly, we find that ISAC is always beneficial, when the duration of generating a sample is more than the duration of uploading a sample. Otherwise, the ISAC gain can vanish or even be negative. Nevertheless, we still derive a sufficient condition, under which a positive ISAC gain is feasible. 
 
\end{abstract}

\begin{IEEEkeywords}
Beamforming, edge intelligence, integrated sensing and communication (ISAC).
\end{IEEEkeywords}

\section{Introduction}
Edge intelligence emerges as a promising solution to leverage massive data distributed at the network edge for training various machine learning models at the edge server \cite{0,1,2}. 
Generally, edge intelligence can be categorized into two types: federated learning and centralized learning. 
If local computing is available at the devices, federated learning can be adopted, where all the devices update and upload their local learning model parameters periodically to the edge server for model training \cite{5,6,7}. 
On the other hand, if local computing is not available, centralized learning is required, where devices need to upload the sensing data to the edge server via wireless communications \cite{11,12,13}.
This is typically the case in sensor networks where the energy and computation resources at the devices are rather limited. In \cite{11}, a learning-centric power allocation was proposed to optimize the generalization performance by smart dataset collection from many sensors. 
In \cite{12}, the path, power, and sample amount planning for unmanned ground vehicles were optimized to maximize the generalization performance in centralized edge intelligence systems. 
In \cite{13}, robotic experiments were carried out to verify the robustness of centralized learning against motion and communication uncertainties.

Including \cite{5,6,7,11,12,13}, enabling edge intelligence systems can be generally divided into four stages: sensing, communication, training, and inference stages. Specifically, it senses the diverse data by sensors, collects the data via communication, trains machine learning models at the edge server, and executes model inference for the mobile users. However, existing edge intelligence systems  \cite{5,6,7,11,12,13} execute the sensing and communication stages sequentially, resulting in excessive amount of dataset generation and uploading time. 
To accelerate edge intelligence, this paper proposes to merge the sensing and communication stages so as to make the best use of the wireless signals for the dual purpose of dataset generation and uploading, so called integrated sensing and communication (ISAC) approach. The ISAC technology has been studied in existing literature \cite{21,23,24}. 
In \cite{21}, the optimal design of dual-functional waveform for radar and communication was proposed. 
In \cite{23}, the reconfigurable intelligent surface-aided dual-functional ISAC system was devised. In \cite{24}, the joint trajectory, sensing, and communication design for unmanned aerial vehicles (UAV) were considered.
Nevertheless, none of the above research have considered to apply ISAC in accelerating sensing and communication stages for edge intelligence.

To fill the research gap, this paper proposes to accelerate the edge intelligence via ISAC.  Our contributions are summarized as follows: We propose the ISAC-assisted edge intelligence system, where the sensor simultaneously senses the target and uploads the samples to the edge server.
To optimize the generalization performance of proposed system, we formulate a non-convex beamforming and time allocation problem. We derive the globally optimal solution of the formulated problem by leveraging the Charnes-Cooper transformation, semidefinite relaxation (SDR), and Karush-Kuhn-Tucker (KKT) conditions. 
We also analyze the computational complexity of proposed solutions. We analyze the ISAC gain over conventional edge intelligence system with sequential sensing and communication stages. Specifically, we reveal that ISAC is always beneficial, when the duration of generating a sample is more than the duration of uploading a sample, and analyze the impact of system parameters on ISAC gain. When the duration of uploading a sample is more than the duration of generating a sample, ISAC gain can vanish or even be negative, due to the excessive interference and power splitting by sensing functionality of ISAC. Nevertheless, we still derive a sufficient condition, under which a positive ISAC gain is feasible.   Furthermore, we demonstrate the effectiveness of the proposed system via a simulated use case on human motion recognition. The simulation results validate the acceleration of proposed system and corroborate our ISAC gain analysis.

\textit{Notations}: $a$ is a scalar, $\textbf{a}$ is a column vector, and $\textbf{A} $ is a matrix. $|\cdot|$ denotes modulus operator. $||\cdot||$ denotes Euclidean norm. $(\cdot)^H$ denotes conjugate transpose operator. 

\textit{Organization}: The remaining paper is organized as follows: We first introduce the system model of ISAC-assisted edge intelligence system in Section-II. Then, we formulate the problem of joint time allocatio and beamforming  optimzation in Section-III. Next, we present the solution approach for global optimal solution and the analysis of ISAC gain in Section-IV and Section-V. The simulation results are presented in Section-VI. Finally, we concluded this paper in Section-VII.

\section{ISAC-assisted Edge Intelligence System Model}

\begin{figure}[t]
	\centering
	\includegraphics[width=0.49\textwidth]{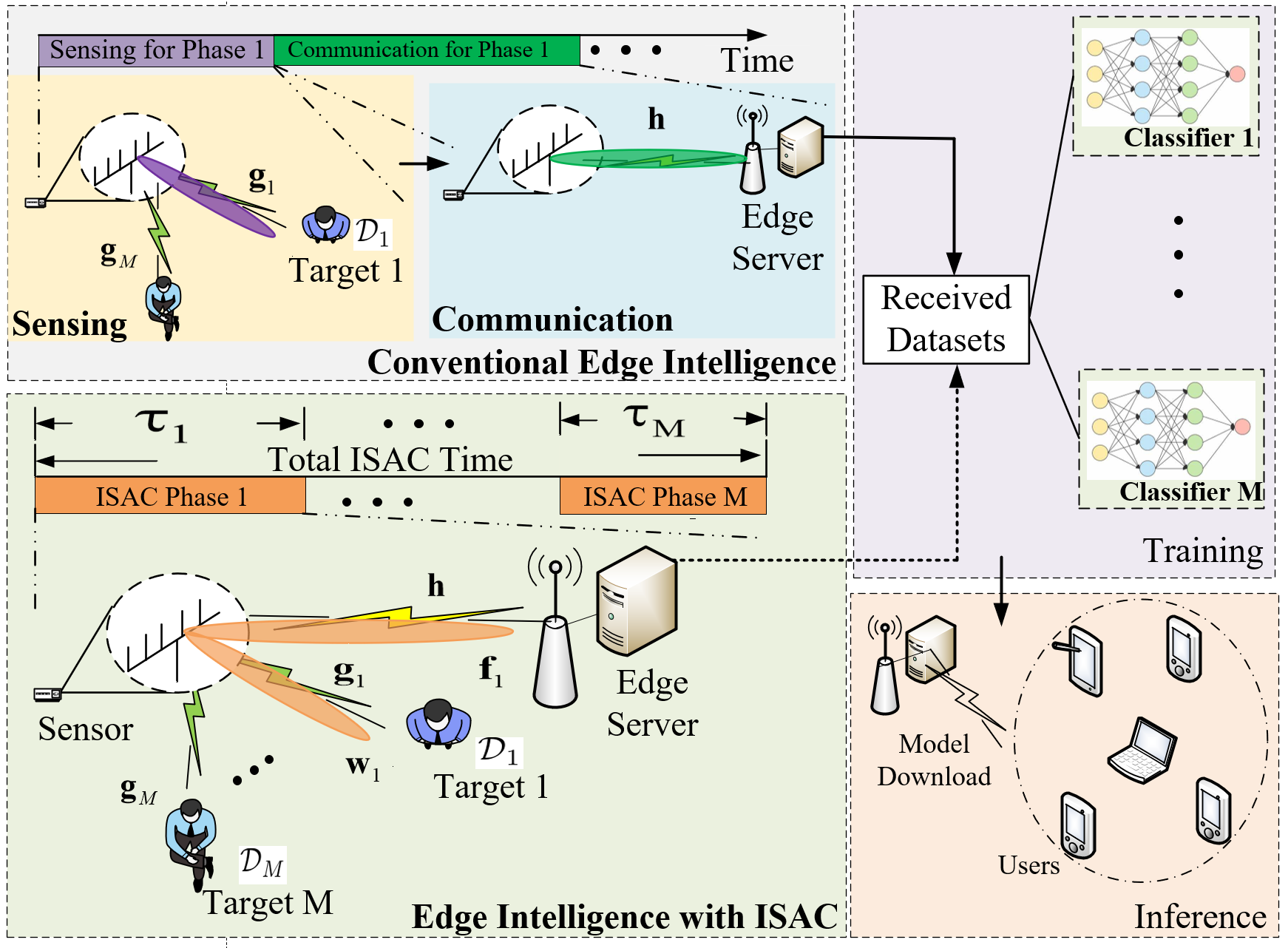}
	\caption{Proposed ISAC-assisted edge intelligence system v.s. conventional edge intelligence system, where the conventional system is with sequential sensing and communication stages, and the proposed edge intelligence system has a merged sensing and communication stage via ISAC.}
	\label{F1}
\end{figure}

We consider an edge intelligence system, consisting of one sensor with $N$ antennas, $M$ sensing targets, and one edge server with single antenna. The sensor collects the training datasets from sensing $M$ targets in a time-division manner, and uploads the datasets to the edge server, where the model training corresponding to each target are executed. In contrast to sequential dataset sensing and uploading, we propose to assist the edge intelligence system with ISAC shown in Fig. 1, where the dataset sensing and uploading are performed in parallel, such that the overall time of dataset sensing and uploading may be reduced, namely accelerating. More specifically, the proposed edge intelligence system contains $M$ ISAC phases, where the aim of the $m$-th Phase ($m=1,2,\cdots,M$) is to the collect dataset $m$ from the target $m$ and uploads. In the $m$-th Phase, there are $L_m$ epoches, where each epoch generates one training sample from sensing the target. The duration of $m$-th Phase and each epoch are denoted by $\tau_m$ and $t_S$, respectively.   

\subsection{Radar Sensing Model}
At the $l$-th epoch of Phase $m$, based on \cite{112}, the radar echo channel between the sensor and the target $m$ is given by
\begin{equation}
	\textbf{G}_{m,l}(t) = g_{m,l}  \underbrace{{\bm{\alpha}}(N,\theta_m){\bm{\alpha}}^T(N,\theta_m)}_{\text{denoted by}\,\hat{\textbf{G}}_m} \delta(t-d_{m,l}),
\end{equation} 
where $g_{m,l}$ denotes the reflection coefficient including the round-trip
pathloss, radar cross section, and Doppler phase shift; ${\bm{\alpha}}(N,\theta_m) \triangleq [1; e^{j2\pi d/ \lambda \sin \theta_m}; \cdots; e^{j(N - 1)2\pi d/ \lambda \sin \theta_m}]$ with $d$ antenna spacing, $\lambda$ wavelength, and $\theta_m$ angle of departure (AoD) and angle of arrival (AoA); and $\delta(t-d_{m,l})$ denotes the impulse function with propagation delay $d_{m,l}$. In the Phase $m$, the radar clutter channel between the sensor and the target $m$ is given by\footnote{Since the beamformer and sensing environment is not changed in the $m$-th Phase, we therefore assume that the radar clutter channel is unchanged for each epoch within the $m$-th Phase.}
\begin{equation}
	\textbf{C}_{m}(t) = \sum_{q=1}^{Q} c_{m,q} 
	{\bm{\alpha}}(N,\psi_{m,q}){\bm{\alpha}}^T(N,\phi_{m,q})\delta(t-d_{m,q}),
\end{equation}
where there are $Q$ multipaths; $c_{m,q}$ denotes the reflection coefficient for the $q$-th multipath; $\psi_{m,q}$ denotes the AoA, and $\phi_{m,q}$ denotes the AoD; $d_{m,q}$ denotes the propagation delay for the $q$-th multipath. At the $l$-th epoch of Phase $m$, the transmit signal from the sensor combines sensing and communication functionalities, which contains radar signal $s_{m,l}^r(t)$, e.g., FMCW, with radar beamformer $\textbf{w}_m$ and communication signal $s_{m,l}^d(t)$ with communication beamformer $\textbf{f}_m$. Thanks to the asymptotic orthogonality of communication beamformer and radar echo channel and the communication signal interference cancellation, we ignore the impact of communication signal when examining the received radar signal\footnote{See \cite{116} and reference therein for self-interference cancellation in ISAC.}. At the $l$-th epoch of Phase $m$, the received radar signal at the sensor is given by 
\begin{eqnarray}
	&& \textbf{y}_{m,l}(t) = (\textbf{G}_{m,l}(t) + \textbf{C}_{m}(t))* s_{m,l}^r(t), \nonumber \\
	&& =  (g_{m,l}   \hat{\textbf{G}}_m  + \underbrace{\sum_{q=1}^{Q} c_{m,q} 
	{\bm{\alpha}}(N,\psi_{m,q}){\bm{\alpha}}^T(N,\phi_{m,q})}_{\text{denoted by}\,\,\textbf{C}_m})\nonumber \\
	&&  \times s^r_{m,l}(t-d_{m,l}) + \textbf{n}_{m,l}(t), \label{S1}
\end{eqnarray}
where $*$ denotes the convolution operator; $\textbf{n}_{m,l}(t)$ denotes the additive white Gaussian noise (AWGN) and follows ${\cal{CN}}(\textbf{0},\sigma^2\textbf{I}_N)$.  Built upon \eqref{S1}, we define the sensing signal-to-noise-plus-interference ratio (SINR) of Phase $m$ as
\begin{equation}
	\textsc{SINR}_m^\text{SEN} = \frac{||\hat{g}_m\hat{\textbf{G}}_m\textbf{w}_m||^2}{\sigma^2 +   ||\textbf{C}_m\textbf{w}_m||^2},
\end{equation}
where $\hat{g}_m$ denotes the amplitude of reflection coefficient with $\hat{g}_m = |g_{m,1}|,=\cdots=|g_{m,L_m}|$, as the distance between the sensor and the target is assumed to be fixed within Phase $m$. All $\hat{g}_m, \hat{\textbf{G}}_m$, and $\textbf{C}_m$ can be estimated before the start of Phase $m$, since distance and angle of target, and environment are assumed to be unchanged during Phase $m$.


\subsection{Communication Model}
At the $l$-th epoch of Phase $m$, based on \cite{112}, the received signal at the edge server is expressed as 
\begin{equation}
	z_{m,l}(t) = \textbf{h}^H \textbf{w}_m s_{m,l}^r(t-p_{m,l}) + \textbf{h}^H \textbf{f}_ms_{m,l}^d(t-p_{m,l}) + n_{m,l}(t),
\end{equation}
where $\textbf{h}$ denotes the channel between sensor and edge server;  $p_{m,l}$ denotes the propagation delay induced by communication channel; and and $n_{m,l}(t)$ denotes the AWGN with $n_{m,l}(t) \sim {\cal{CN}}(0,\sigma^2)$. The  transmit power constraint is given by
\begin{equation}
	\| \textbf{w}_m\|^2 + \| \textbf{f}_m\|^2 \le P, \qquad \forall m,\label{PowerC}
\end{equation}
where $P$ denotes the maximal transmit power. As such, the communication SINR  at the edge server is defined as
\begin{equation}
	\textsc{SINR}_m^\text{COM} = \frac{|\textbf{h}^H\textbf{f}_m|^2}{\sigma^2 +  \gamma|\textbf{h}^H\textbf{w}_m|^2},
\end{equation}
where $0 \le \gamma \le 1$ is the successive interference cancellation (SIC) coefficient with $1$ for no SIC and $0$ for perfect SIC.

\section{Problem Formulation}

The goal of edge intelligence system is to optimize generalization performance. To this end, the widely-used empirical model for classification error when inference, (e.g., in \cite{11}), is taken into use, given by $a_mv_m^{-b_m}$, where $a_m,b_m > 0$ are hyper-parameters for difficulty of task $m$, and $v_m$ is the number of collected samples of task $m$. The number of collected samples need to satisfy the ISAC constraint. That is, $v_m$ should be equal to the minimum of the number of samples that the communication link can afford and the number of samples that the sensing can provide, given by
\begin{eqnarray}
&& \!\!\!\!\!\!\!\!\!\!\!\!\!\!\!\!\!\! v_m = \nonumber \\ 
&& \!\!\!\!\!\!\!\!\!\!\!\!\!\!\!\!\!\! \min\left\{ \frac{B\tau_m}{D} \log_2 \left(1 +\frac{|\textbf{h}^H\textbf{f}_m|^2}{\sigma^2 + \gamma|\textbf{h}^H\textbf{w}_m|^2} \right), \frac{\tau_m}{t_S} \right\}, \quad 
\forall m,
\end{eqnarray}
where the number of samples that the communication link can afford is given by $\frac{B\tau_m}{D} \log_2 \left(1 +\frac{|\textbf{h}^H\textbf{f}_m|^2}{\sigma^2 + \gamma|\textbf{h}^H\textbf{w}_m|^2} \right)$ with $\tau_m$ duration of Phase $m$, $B$ bandwidth, $D$ data volume per sample, and the number of sensing samples of Phase $m$ is expressed as $\tau_m/t_S$. On the other hand, the corresponding sensing quality constraint should be met, namely the sensing SINR of Phase $m$ is not less than the threshold $\eta_m$, given by 
\begin{equation}
\eta_m \le \frac{||\hat{g}_m\hat{\textbf{G}}_m\textbf{w}_m||^2}{\sigma^2 +   ||\textbf{C}_m\textbf{w}_m||^2}, \qquad \forall m. 
\end{equation}
In addition, the duration of $M$ ISAC phases should be constrained, given by 
\begin{equation}
\sum_{m=1}^M \tau_m \le T. \label{TimeC}	
\end{equation}

Therefore, considering the overall performance and fairness, we aim to minimize the maximal classification error of $M$ tasks w.r.t. beamformers and time allocation for $M$ phases, under four constraints, namely the ISAC constraint, the sensing quality constraint, the duration of $M$ ISAC phases constraint, and the maximal transmit power constraint. Mathematically, the resulting problem is formulated below 
\begin{subequations}
	\begin{eqnarray}
		&& \text{(P1)}: \min_{\{\textbf{w}_m,\textbf{f}_m, {{\tau}_m}\}} \max_{m} \,\,\, a_m v_m^{-b_m} \nonumber \\
		&& \qquad \qquad \qquad \qquad \text{s.t.}   \,\,\, (6),(8)-(10).	\nonumber
	\end{eqnarray}	
\end{subequations}

\textbf{Remark 1}: Based on each author's knowledge, it is the first time that minimizing the maximal  classification error is considered, rather than optimizing the sensing or communication metrics in the majority of existing ISAC work. On the other hand, the existing edge intelligence work omit the benefit of ISAC, which is the main focus of our work. Therefore, the formulation of Problem (P1) is novel, especially the introduction of the ISAC constraint for edge intelligence system. Unfortunately, the Problem (P1) is non-convex. Nevertheless, we still develop the algorithm for globally optimal solution.

\section{Globally Optimal Solution}


The globally optimal solution of Problem (P1) is obtained via two steps. In Step-I, the beamforming is globally optimized via Charnes-Cooper transformation and rank-1 guaranteed SDR, where the transformed problem is not related to time allocation. In Step-II, given the optimal beamformer, the time allocation problem is reformulated as a convex problem and solved by KKT conditions. The details are elaborated below.

\subsection{Step-I: Beamforming Optimization}

Firstly, we shall optimize the beamformers given time allocation.  The resulting problem can be written as 
\begin{subequations}
	\begin{eqnarray}
		&&  \!\!\!\!\!\!  \min_{\{\textbf{w}_m,\textbf{f}_m\}}\max_{m} \,  a_m\left[\frac{B\tau_m}{D} \log_2 \left(1 +\frac{|\textbf{h}^H\textbf{f}_m|^2}{\sigma^2 + \gamma |\textbf{h}^H\textbf{w}_m|^2} \right)\right]^{-b_m} \nonumber \\
		&& \!\!\!\!\!\!  \qquad \qquad \text{s.t.} \,\,\,  \eta_m \le \frac{||\hat{g}_m\hat{\textbf{G}}_m\textbf{w}_m||^2}{\sigma^2 +   ||\textbf{C}_m\textbf{w}_m||^2}, \qquad  \forall m,  \label{2Q} \\		
		&& \!\!\!\!\!\!  \qquad \qquad \qquad ||\textbf{w}_m||^2 + ||\textbf{f}_m||^2 \le P,  \qquad \forall m,  \label{4Q}
	\end{eqnarray}	
\end{subequations}
Letting $\textbf{W}_m = \textbf{w}_m\textbf{w}^H_m, \, \textbf{F}_m = \textbf{f}_m\textbf{f}_m^H, \, \textbf{H} = \textbf{h}\textbf{h}^H$, $\textbf{D}_m = |\hat{g}_m|^2\hat{\textbf{G}}_m^H\hat{\textbf{G}}_m$, $\textbf{E}_m = \textbf{C}_m^H\textbf{C}_m$, $\textbf{W}'_m = \textbf{W}_m\xi, \,\textbf{F}'_m = \textbf{F}_m\xi$, and $\xi > 0$, we can decouple the problem (12) into $M$ semidefinite programming (SDP) subproblems. The $m^{th}$ subproblem is
\begin{subequations}
	\begin{eqnarray}
		\max_{\{\textbf{W}_m', \textbf{F}_m'\},\xi} && \!\!\!\!\!  \text{Tr}\{\textbf{F}_m'\textbf{H}\} \nonumber \\
		\text{s.t.} && \!\!\!\!\! \sigma^2 \xi + \gamma\text{Tr}\{\textbf{W}_m'\textbf{H}\} = 1,  \\
		&& \!\!\!\!\!  \eta_m \sigma^2 \xi + \text{Tr}\{\textbf{E}_m\textbf{W}_m'\}   \le  \text{Tr}\{\textbf{D}_m\textbf{W}_m'\}, \\  
		&& \!\!\!\!\!  \text{Tr}\{\textbf{W}_m'\} + \text{Tr}\{\textbf{F}_m'\} \le \xi P, \label{1QQQQQ} \\
		&&  \!\!\!\!\! \textbf{W}_m',\textbf{F}_m' \succeq \textbf{0}. 
	\end{eqnarray}	
\end{subequations}

\textbf{Proposition 1}: The Problem (13) has the same optimal solution as that of Problem (12). 

\begin{IEEEproof}
Please refer to Appendix A.
	\end{IEEEproof}

The above SDP problem can be solved by the CVX toolbox. Therefore, the Problem (12) can be solved as well.

\subsection{Step-II: Time Allocation Optimization}

Secondly, we shall optimize the time allocation given optimal beamformers. The resulting problem can be written as 
		\begin{eqnarray}
		 &&  \!\!\!\!\!\! \min_{\{{{\tau}_m}\}} \max_{m} \,  a_m\left[\min\left\{\frac{B\tau_m}{D} \log_2 \left(1 +\frac{|\textbf{h}^H\textbf{f}^*_m|^2}{\sigma^2 + \gamma|\textbf{h}^H\textbf{w}^*_m|^2} \right),\right.\right. \nonumber \\
		&& \!\!\!\!\!\!  \qquad \qquad \left.\left.\frac{\tau_m}{t_S}\right\}\right]^{-b_m}  \nonumber\\	 	
		&& \!\!\!\!\! \qquad \quad \text{s.t.} \,\,\, \sum_{m=1}^M \tau_m \le T. \label{1W}  
	\end{eqnarray}

\textbf{Proposition 2}: The optimal solutions of Problem (14), denoted by $\tau_m^*,\,\forall m$,  satisfy the following optimality conditions: 
\begin{equation}		
	\tau_m^* = \frac{1}{\min\{\pi_{m,c}(\text{SINR}_m^\text{COM}),\pi_{s}(t_S) \}}\left(\frac{\mu^*}{a_m}\right)^{\frac{-1}{b_m}} ,\,\, \forall m, \label{K6}
\end{equation}
where $\mu^*$ satisfies
$
	\sum_{m=1}^M \frac{1}{\min\{\pi_{m,c}(\text{SINR}_m^\text{COM}),\pi_{s}(t_S) \}}\left(\mu^*/a_m\right)^{\frac{-1}{b_m}}$ $= T$,  
 $ \pi_{m,c}(\text{SINR}_m^\text{COM}) = B \log_2 \left(1 + \text{SINR}_m^\text{COM} \right)/D$ with $\text{SINR}_m^\text{COM} = |\textbf{h}^H\textbf{f}^*_m|^2/(\sigma^2 + \gamma|\textbf{h}^H\textbf{w}^*_m|^2)$, and $\pi_{s}(t_S) = 1/t_S$.

\begin{IEEEproof}
 Please refer to Appendix B for the proof. 
\end{IEEEproof}

\subsection{Overall Algorithm and Complexity Analysis}

Now, we  finally summarize the proposed approach for solving Problem (P1) in Algorithm 1. The  worst case complexity of Algorithm 1 is analyzed below. In  Step-I, the complexity of solving Problem (13) via SDR is ${\cal{O}}(N^{3.5})$ \cite{100}. Since we need to address $M$ problems, the total complexity of Step-I is ${\cal{O}}(MN^{3.5})$. In  Step-II, the complexity of bisection search for $\mu^*$ is ${\cal{O}}(M\log(1/\epsilon))$ \cite{105}, as having $M$ components. The total complexity of Step-II is ${\cal{O}}(M+M\log(1/\epsilon))$, due to \eqref{K6} and bisection search. Overall, the complexity of proposed algorithm is polynomial, i.e.,  
${\cal{O}}(M+M\log(1/\epsilon) + MN^{3.5})$.
\begin{algorithm}[t]
	\caption{The proposed algorithm for solving Problem (P1)}  
	\hspace*{0.02in} {\bf Input}:   $\sigma^2, B, D, T, t_S, \textbf{h}, \hat{g}_m, \hat{\textbf{G}}_m, \textbf{C}_m$, $a_m, b_m, \eta_m,\,\forall m$
	\begin{algorithmic}[1]
		\State For $m=1:M$
		\State \quad Solving Problem (13) by CVX. \% \textit{Subproblem}
		\State End
		\State Eigendecomposition of optimal solution of Problem (13).
		\State Solving Problem (14) via bisection search  and \eqref{K6}.
	\end{algorithmic}
	\hspace*{0.02in} {\bf Output}: $\textbf{w}_m^*,\,\textbf{f}_m^*,\,{\tau}_m^*,\,\forall m.$
\end{algorithm}

\section{ISAC Gain Analysis}

\subsubsection{Sensing Dominance} 
When the duration of generating a sample is more than the duration of uploading a sample (i.e., sensing dominance), the ISAC gain over that of conventional edge intelligence systems is defined and analyzed below.
\begin{eqnarray}	
		&& \!\!\!\!\!\!\!\!\! \textbf{\textit{ISAC Gain}} = 1 - \frac{\text{ISAC Time}}{\text{Conventional Sens. \& Com. Time}} \nonumber \\
		&& \!\!\!\!\!\!\!\!\!  =    \dfrac{\sum_{m=1}^M v_mD/ (B  \log_2(1+P||\textbf{h}||^2/\sigma^2))}{\sum_{m=1}^M v_mt_S + \sum_{m=1}^M v_mD/ (B \log_2(1+P||\textbf{h}||^2/\sigma^2))} \nonumber\\
		&& \!\!\!\!\!\!\!\!\!  = \dfrac{1}{t_S B\log_2(1+P||\textbf{h}||^2/\sigma^2)/D+1} > 0,  \label{GISAC}
\end{eqnarray}
where the proposed ISAC-assisted and conventional edge intelligence systems have the same amount of training samples, and the maximal ratio transmit (MRT) beamforming is taken for the conventional edge intelligence systems. 

\textbf{Remark 2}: The ISAC gain analysis in \eqref{GISAC} exhibits: 1) The ISAC gain is always positive when sensing dominance; and 2) The ISAC gain is inversely proportional to the sensing time per sample and data volume per sample, and proportional to bandwidth and maximal transmit power.
 
\subsubsection{Communication Dominance} 
When the duration of uploading a sample is more than the duration of generating a sample   (i.e., communication dominance), due to the excessive interference and power splitting by sensing functionality of ISAC,  ISAC gain can vanish or even be negative. Nevertheless, a positive ISAC gain is possible, if a sufficient condition holds, which is shown below.  
\begin{eqnarray}	
	&& \!\!\!\!\!\!\!\!\! \textbf{\textit{ISAC Gain}} =1 - \frac{\text{ISAC Time}}{\text{Conventional Sens. \& Com. Time}}\nonumber \\
	&& \!\!\!\!\!\!\!\!\! \overset{(a)}{=} \dfrac{\sum_{m=1}^M v_mt_S + \sum_{m=1}^M v_mD/ (B  \log_2(1+P||\textbf{h}||^2/\sigma^2))}{T_\text{Conv.}} \nonumber\\
	&& \!\!\!\!\!\!\!\!\! \quad - \dfrac{\sum_{m=1}^M v_mD/ (B  \log_2(1+|\textbf{h}^H\textbf{f}_m|^2/\sigma^2))}{T_\text{Conv.}} \nonumber\\
		&& \!\!\!\!\!\!\!\!\! \overset{(b)}{=}    \dfrac{\sum_{m=1}^M v_mt_S + \sum_{m=1}^M v_mD/ (B  \log_2(1+ P||\textbf{h}||^2/\sigma^2))}{T_\text{Conv.}} \nonumber\\
	&& \!\!\!\!\!\!\!\!\! \quad - \dfrac{\sum_{m=1}^M v_mD/ (B  \log_2(1+\beta P||\textbf{h}||^2/\sigma^2))}{T_\text{Conv.}} \nonumber\\
	&& \!\!\!\!\!\!\!\!\! \overset{(c)}{\ge}   \dfrac{\sum_{m=1}^M v_mD/ (\beta B  \log_2(1+P||\textbf{h}||^2/\sigma^2))}{T_\text{Conv.}} \nonumber\\
		&& \!\!\!\!\!\!\!\!\! \quad  - \dfrac{\sum_{m=1}^M v_mD/ (B  \log_2(1+\beta P||\textbf{h}||^2/\sigma^2))}{T_\text{Conv.}}  \overset{(d)}{>} 0,  
	\label{CGISAC}
\end{eqnarray}
where (a) is from $\gamma = 0$ and $T_\text{Conv.} = \sum_{m=1}^M v_mt_S + \sum_{m=1}^M v_mD/ (B \log_2(1+P||\textbf{h}||^2/\sigma^2)$; (b) is from letting $\textbf{f}_m = \sqrt{\beta P} \textbf{h}, \,\forall m$ with power splitting factor $\beta,\,0 \le \beta \le  1$; (c) is from letting $t_S \ge (\frac{1}{\beta}-1)D/ (B  \log_2(1+P||\textbf{h}||^2/\sigma^2))$; (d) is from Jensen's inequality. Moreover, the analysis in \eqref{CGISAC} implies that there exists a threshold of $\eta_m$, denoted by $\eta_m'$, above which the positive ISAC gain is feasible. Specifically, due to $\gamma = 0$, $\textbf{f}_m = \sqrt{\beta P} \textbf{h}, \,\forall m$, and power splitting from maximal transmit power $P$, the radar beamformer $\textbf{w}_m$ can be designed as $\sqrt{(1-\beta)P}\textbf{w}_m'$, where $\textbf{w}_m'$ is given by 
\begin{equation}
\textbf{w}_m'	= \text{argmax}_{\|\textbf{w}_m\|_2^2 = 1}\frac{(1-\beta)P||\hat{g}_m\hat{\textbf{G}}_m\textbf{w}_m||^2}{\sigma^2 +   (1-\beta)P||\textbf{C}_m\textbf{w}_m||^2}.
\end{equation}
It can be seen that $\eta_m'$ needs to satisfy
\begin{equation}
	\eta_m'	\le   \frac{(1-\beta)P||\hat{g}_m\hat{\textbf{G}}_m\textbf{w}'_m||^2}{\sigma^2 +   (1-\beta)P||\textbf{C}_m\textbf{w}'_m||^2}. \label{ETA}
\end{equation}

Therefore, we are able to conclude that if a sufficient condition holds, that is, $\gamma = 0$, $0 \le \beta \le 1$, $\textbf{f}_m = \sqrt{\beta P} \textbf{h}, \,\forall m$, $t_S \ge (\frac{1}{\beta}-1)D/ (B  \log_2(1+P||\textbf{h}||^2/\sigma^2))$, and $\{\eta_m', \,\forall m\}$ satisfying \eqref{ETA}, a positive ISAC gain is feasible.

   \begin{figure}[t]
	\centering
	\includegraphics[width=3.45in]{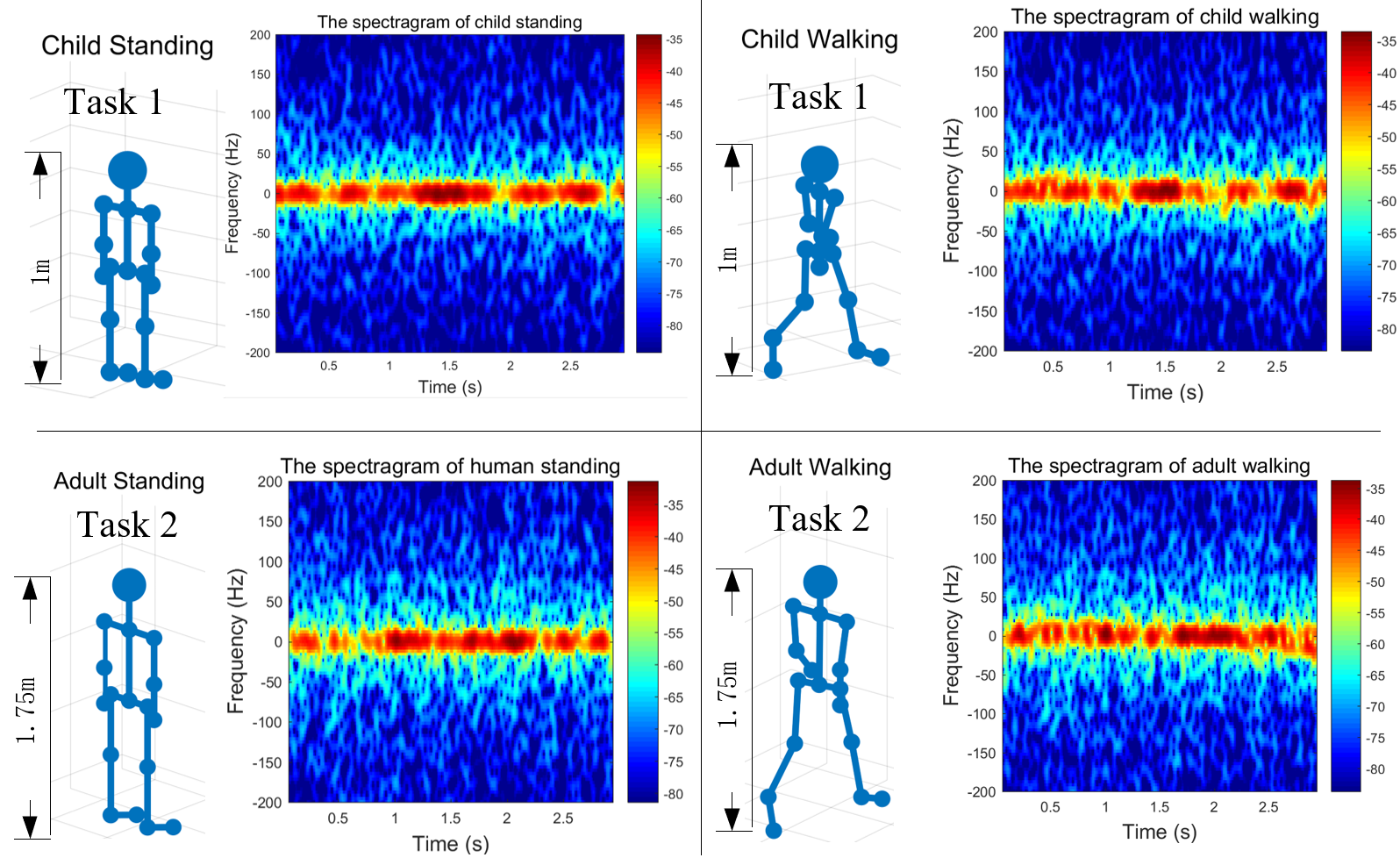}
	\caption{Generated human motion and spectrogram from platform \cite{106}.} \label{Fig4}
\end{figure}

\textbf{Remark 3}: The above sufficient condition shows that with perfect SIC,  large enough sensing time, properly selected beamformers, and qualified threshold of sensing quality, a positive ISAC gain is feasible when communication dominance.

\section{Simulations}

%

\begin{figure}
			\centering
	\includegraphics[width=2.5in]{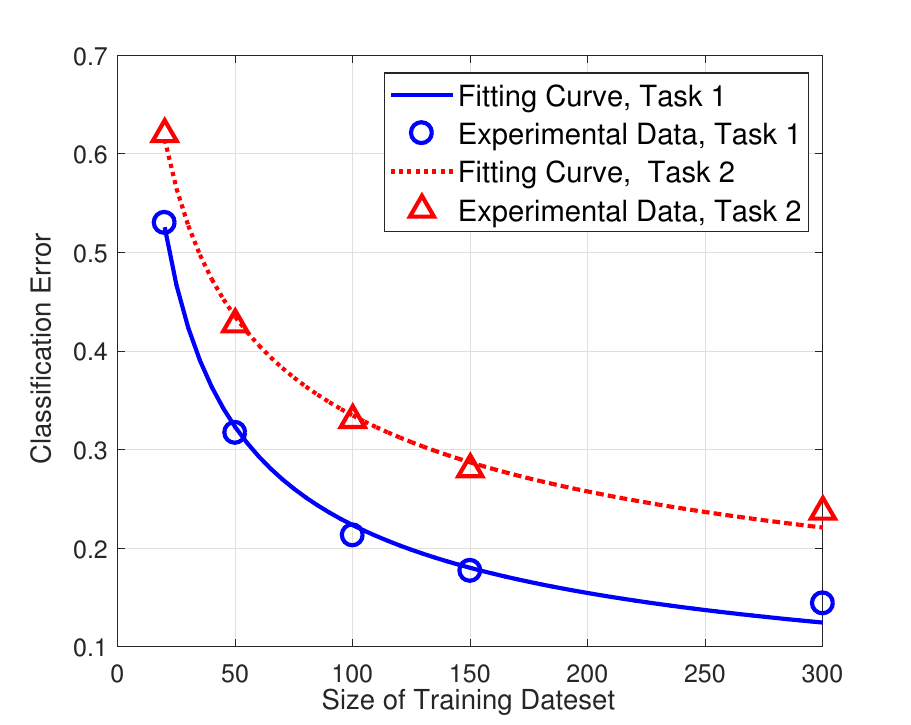}
	\caption{Experimental curve fitting.} \label{Fig5}
			\centering
	\includegraphics[width=2.5in]{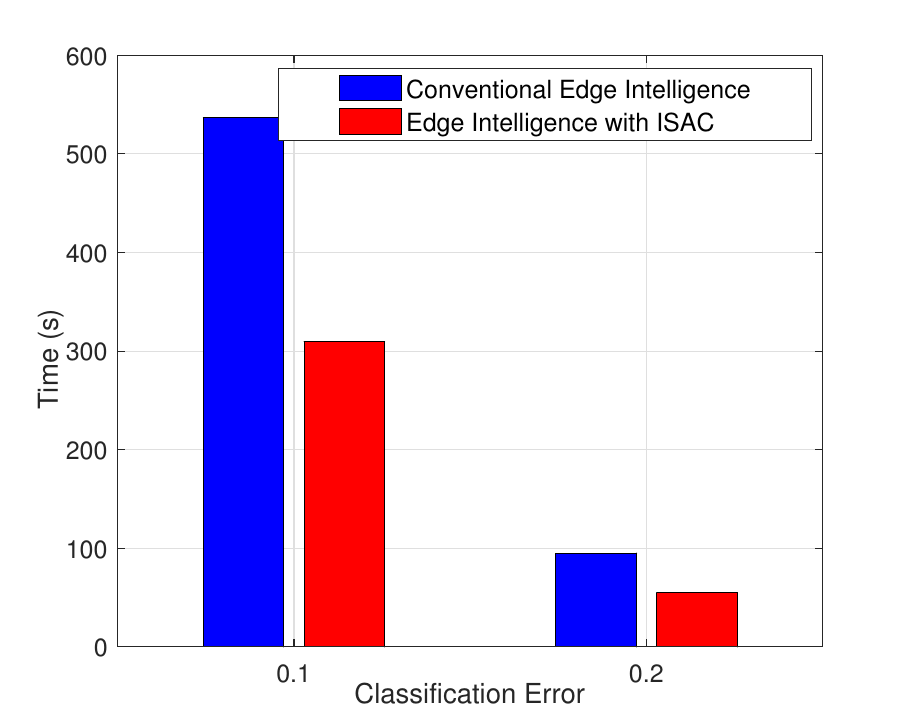}
	\caption{Time-saving by ISAC.} \label{Fig6}
\end{figure}

\textbf{Human Motion Recognition Tasks}: The generated spectrograms via human motion recognition platform \cite{106} are shown in Fig. \ref{Fig4}. In particular, there are 2 targets, i.e., 1 child with $1$m height and 1 adult with $1.75$m, where the child and adult are located at $90^0$, $20$m, and $270^0$, $40$m  from the sensor, respectively. The edge server is $250$m away from the sensor. To classify the human motion, we build two CNNs, where hidden layers are set as 32, 64, and 128 units for each CNN.   
Correspondingly, there are 2 tasks, where task 1 is to classify the motion of child standing and walking, and task 2 is to classify the motion of adult standing and walking.  In Fig. \ref{Fig5}, we fit the experimental data to obtain the hyper-parameters of fitting curve, where $(a_1, b_1) = (2.5845, 0.5317)$ and $(a_2, b_2)=(1.9057, 0.3778)$. Moreover, the remaining simulation parameters are set as follows: As $Q$ is large,  $\textbf{C}$ can be approximated by complex Gaussian matrix, whose item follows ${\cal{CN}}(0,-70\text{dBm})$. The large-scale fading factor is $2.5$. The number of antennas at sensor is 4. Antenna spacing is $0.15$m. Wavelength is $0.3$m. SIC coefficient at edge server is $1$. The maximal transmit power is $1$W. Data volume per sample is $1$Mbits. Bandwidth is $5$MHz. AWGN power is $-90$dBm. Sensing threshold for task 1 and 2 are $20$dB and $1$dB, respectively. Simulation results are given below.


\textbf{Time-saving by ISAC}: In Fig. \ref{Fig6}, we evaluate the time-saving of proposed ISAC-assisted system over conventional system, where $t_S = 0.1s$, and MRT beamforming is adopted for conventional system. Fig. \ref{Fig6} shows that ISAC accelerates the stages of sensing and communication, where the time-saving by ISAC becomes less as classification error increases. The reason is that the size of training datasets becomes smaller as classification error increases, thus reducing the amount of time saved by  communication and sensing stage merging.  

\begin{figure}
			\centering
	\includegraphics[width=2.5in]{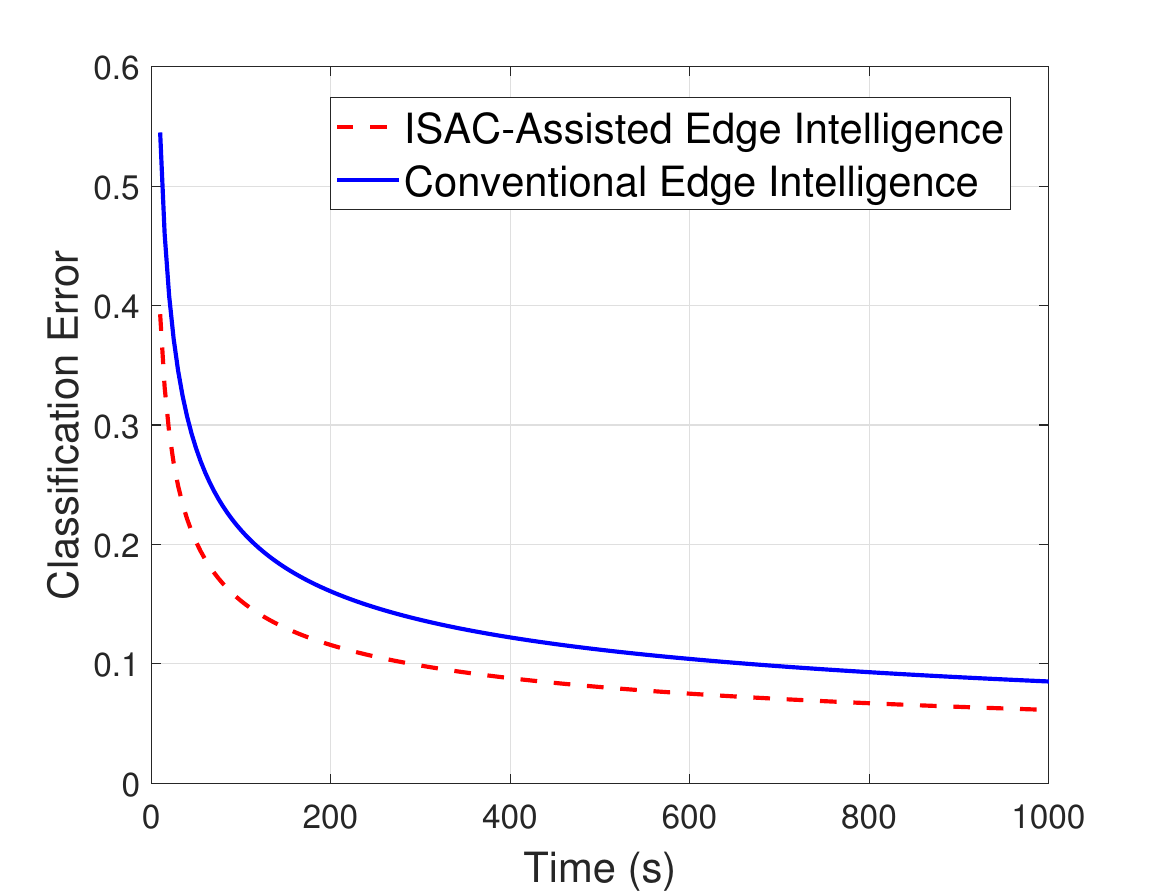}
	\caption{Error reduction by ISAC.}
	\label{Fig8}
			\centering
	\includegraphics[width=2.5in]{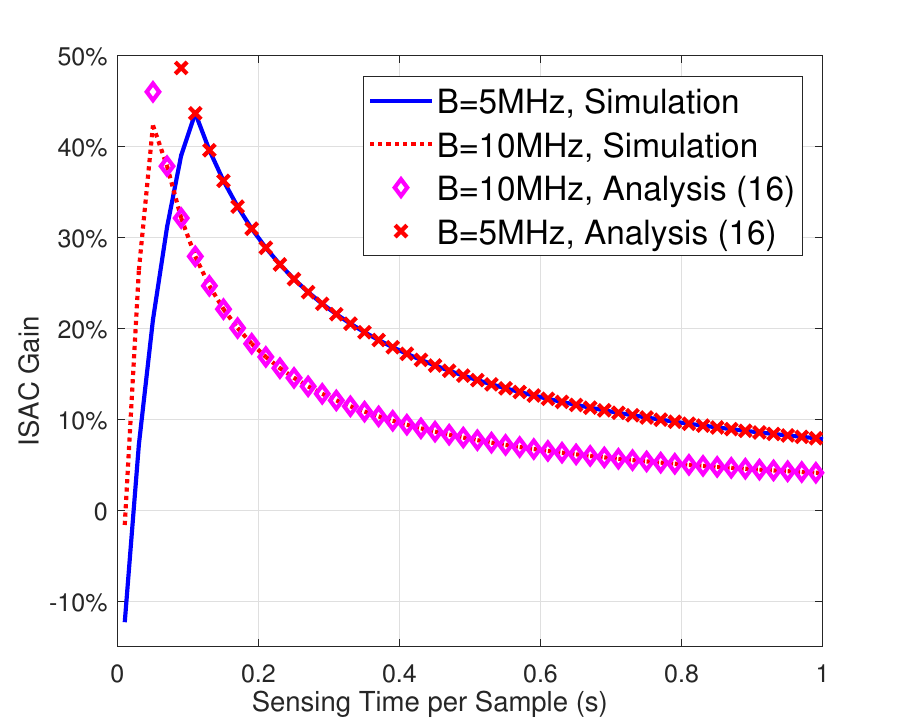}
	\caption{Impacts on ISAC gain.}
	\label{Fig7}
\end{figure}

\textbf{Error Reduction by ISAC}: In Fig. \ref{Fig8}, we evaluate the classification error reduction by employing the proposed ISAC-assisted system, where $t_S = 0.1s$, and MRT beamforming is adopted for conventional system. With the same amount of time, the proposed ISAC-assisted system enjoys an overall classification error reduction. This error reduction exhibits that employing ISAC can collect more samples by using the same amount of time, in contrast to conventional edge intelligence.

\textbf{Impacts on ISAC Gain}: In Fig. \ref{Fig7}, we evaluate the ISAC gain of proposed ISAC-assisted system over conventional system.  Fig. \ref{Fig7} shows that the ISAC gain first increases from a negative value to the maximal value, then gradually decreases, as  $t_S$ increases. Fig. \ref{Fig7} shows that our analysis of ISAC gain \eqref{GISAC} matches with simulation results for sensing dominance, and validates that the ISAC gain is inversely proportional to $t_S$ in this case. In addition, it shows that a larger bandwidth leads to a smaller ISAC gain for sensing dominance, corroborating our ISAC gain analysis. On the other hand, when communication dominance, Fig. \ref{Fig7} shows that the ISAC gain can vanish or even be negative. This is because the excessive interference and power splitting by sensing functionality of ISAC. 






\section{Conclusions}

We proposed the ISAC-assisted edge intelligence system, where the sensing and communication stages were merged so as to make the best use of the wireless signals for the dual purpose of dataset generation and uploading.  Both simulation results and performance analysis show that the ISAC-assisted edge intelligence system is better than the conventional edge intelligence system without ISAC, when sensing dominance. Otherwise, when communication dominance, employing ISAC for edge intelligence system may not be beneficial in terms of saving sample generating and uploading time, due to the excessive interference and power splitting. Nevertheless, for communication dominance, we still identified a sufficient condition, under which a positive ISAC gain is feasible.

\bibliographystyle{IEEEtran}
\bibliography{SDGTRef}


\begin{appendices}
 	
 	\section{Proof of Proposition 1}
 	It can be seen that Problem (12) is decomposable. Moreover, the objective  function is monotonically decreasing w.r.t. $\frac{|\textbf{h}^H\textbf{f}_m|^2}{\sigma^2 + \gamma|\textbf{h}^H\textbf{w}_m|^2}$. As such, it suffices to tackle $M$ separated subproblems, where the $m$-th sub-problem is given by
 	\begin{subequations}
 		\begin{eqnarray}
 			\max_{\{\textbf{w}_m,\textbf{f}_m\}} &&  \frac{|\textbf{h}^H\textbf{f}_m|^2}{\sigma^2 + \gamma|\textbf{h}^H\textbf{w}_m|^2} \nonumber \\
 			\text{s.t.} &&  \eta_m \le \frac{||\hat{g}_m\hat{\textbf{G}}_m\textbf{w}_m||^2}{ \sigma^2 +  ||\textbf{C}_m\textbf{w}_m||^2 },    \label{2QQ} \\	 
 			&& ||\textbf{w}_m||^2 + ||\textbf{f}_m||^2 \le P,     \label{4QQ}
 		\end{eqnarray}	
 	\end{subequations}
 	Applying SDR to tackle the Problem (20),
 	we define $\textbf{W}_m = \textbf{w}_m\textbf{w}^H_m, \, \textbf{F}_m = \textbf{f}_m\textbf{f}_m^H, \, \textbf{H} = \textbf{h}\textbf{h}^H$, $\textbf{D}_m = |\hat{g}_m|^2\hat{\textbf{G}}^H_m\hat{\textbf{G}}_m$, $\textbf{E}_m = \textbf{C}_m^H\textbf{C}_m$, and drop the non-convex rank-$1$ constraint. The relaxed problem of Problem (20) is given by 
 	\begin{subequations}
 		\begin{eqnarray}
 			\max_{\{\textbf{W}_m,\textbf{F}_m\}} &&  \frac{\text{Tr}\{\textbf{F}_m\textbf{H}\}}{\sigma^2 + \gamma \text{Tr}\{\textbf{H}\textbf{W}_m\}} \nonumber \\
 			\text{s.t.} &&  \eta_m \le \frac{\text{Tr}\{\textbf{D}_m\textbf{W}_m\}}{\sigma^2 +   \text{Tr}\{\textbf{E}_m\textbf{W}_m\}},    \label{2QQQ} \\	 
 			&& \text{Tr}\{\textbf{W}_m\} + \text{Tr}\{\textbf{F}_m\} \le P, \\
 			&&  \textbf{W}_m,\textbf{F}_m \succeq \textbf{0},    \label{4QQQ}
 		\end{eqnarray}	
 	\end{subequations}
 	which is a quasi-convex problem, due to \eqref{2QQQ}.
 	Next, we apply the Charnes-Cooper transformation \cite{104} to deal with the quasi-convexity. Specifically, defining $\textbf{W}'_m = \textbf{W}_m\xi, \,\textbf{F}'_m = \textbf{F}_m\xi$, and $\xi > 0$, we reformulate the Problem (21) as  a SDP problem
 	\begin{subequations}
 		\begin{eqnarray}
 			\max_{\{\textbf{W}_m', \textbf{F}_m'\},\xi} && \!\!\!\!\!\!\! \text{Tr}\{\textbf{F}_m'\textbf{H}\} \nonumber \\
 			\text{s.t.} && \!\!\!\!\!\!\! \sigma^2 \xi + \gamma \text{Tr}\{\textbf{H}\textbf{W}_m'\} = 1,  \\
 			&& \!\!\!\!\!\!\! \eta_m \sigma^2\xi + \text{Tr}\{\textbf{E}_m\textbf{W}'_m\}    \le  \text{Tr}\{\textbf{D}_m\textbf{W}_m'\}, \\ 
 			&& \!\!\!\!\!\!\! \text{Tr}\{\textbf{W}_m'\} + \text{Tr}\{\textbf{F}_m'\} \le \xi P, \label{1QQQQ} \\
 			&& \!\!\!\!\!\!\! \textbf{W}_m',\textbf{F}_m' \succeq \textbf{0}, 
 		\end{eqnarray}	
 	\end{subequations}
 	which can be solved by the CVX toolbox. Note that  $\xi > 0$ is implicitly guaranteed by the constraint \eqref{1QQQQ}.  The optimal solution of Problem (22) is obtained by the transform $\textbf{W}_m^* = \textbf{W}_m'^*/\xi^*, \, \textbf{F}_m^* = \textbf{F}_m'^*/\xi^*$, where the optimal solution of Problem (21) is denoted by $\textbf{W}_m'^*,\,\textbf{F}_m'^*,\,\xi^*$. Moreover, the SDR globally optimally solves Problem (20). Namely assuming that $\textbf{W}_m'^*,\, \textbf{F}_m'^* \ne \textbf{0}$, we have  $\text{rank}\{\textbf{W}_m'^*\} = 1$ and $\text{rank}\{\textbf{F}_m'^*\} = 1$.  
 	The rationale is given as follows: Denoting the optimal objective value of Problem (23) as $v_\text{opt}$, we have  
 	\begin{subequations}
 		\begin{eqnarray}
 			\min_{\{\textbf{W}_m', \textbf{F}_m'\},\xi} &&   \!\!\!\!\!\!\!   \text{Tr}\{\textbf{W}_m'\} + \text{Tr}\{\textbf{F}_m'\} - \xi P\nonumber \\
 			\text{s.t.} && \!\!\!\!\!\!\! \sigma^2 \xi + \gamma \text{Tr}\{\textbf{W}_m'\textbf{H}\} = 1,  \\
 			&& \!\!\!\!\!\!\! \eta_m \sigma^2\xi + \text{Tr}\{\textbf{E}_m\textbf{W}'_m\}   \le  \text{Tr}\{\textbf{D}_m\textbf{W}_m'\},\\		 
 			&& \!\!\!\!\!\!\! \text{Tr}\{\textbf{F}_m'\textbf{H}\} \ge v_\text{opt}, \label{infeasible}\\
 			&& \!\!\!\!\!\!\! \textbf{W}_m',\textbf{F}_m' \succeq \textbf{0}, 
 		\end{eqnarray}	
 	\end{subequations}
 	which has the same optimal solution of Problem (22). Otherwise, there is a better solution than $\textbf{W}_m'^*,\,\textbf{F}_m'^*$ that achieves a higher value of $\text{Tr}\{\textbf{F}_m'^*\textbf{H}\}$ than $v_\text{opt}$. However, this is impossible, since we can decrease the objective value of (23) until the equality of \eqref{infeasible} is reached. According to \cite[Proposition 3]{102}, since there are three constraints in (23), the optimal solution of Problem (23) follows:
 	\begin{equation}
 		\text{rank}\{\textbf{W}_m'^*\} + \text{rank}\{\textbf{F}_m'^*\} + \text{rank}\{\xi^*\} \le 3.
 	\end{equation}
 	Due to $\textbf{W}_m'^*, \textbf{F}_m'^* \ne \textbf{0}$, we have $\text{rank}\{\textbf{W}_m'^*\} = 1$ and $\text{rank}\{\textbf{F}_m'^*\} = 1$.  This implies that the rank-1 constraint maintains even if it is removed in the SDP problem.

\section{Proof of Proposition 2}
	Representing the maximal classification error of $M$ tasks by a slack variable $\mu$, we reformulate Problem (14) as  
\begin{subequations}
	\begin{eqnarray}
		\min_{\{\tau_m\},\mu} &&  \!\!\!\!\!\!\!\! \mu \nonumber \\
		\text{s.t.} && \!\!\!\!\!\!\!\!  \frac{\left(\frac{\mu}{a_m}\right)^{\frac{-1}{b_m}}}{\min\{\pi_{m,c}(\text{SINR}_m^\text{COM}),\pi_{s}(t_S) \}} \le \tau_m, \, \forall m,   \\
		&& \!\!\!\!\!\!\!\! \sum_{m=1}^M \tau_m \le T,  
	\end{eqnarray}	
\end{subequations}
where \begin{equation}
	\pi_{m,c}(\text{SINR}_m^\text{COM}) = B \log_2 \left(1 + \text{SINR}_m^\text{COM} \right)/D \nonumber 
\end{equation} with $\text{SINR}_m^\text{COM} = |\textbf{h}^H\textbf{f}^*_m|^2/(\sigma^2 + \gamma|\textbf{h}^H\textbf{w}^*_m|^2)$, $\pi_{s}(t_S) = 1/t_S$. Next, we apply KKT conditions on Problem (25).
By introducing the Lagrangian multiples $\lambda_1,\lambda_2,\cdots,\lambda_{M+1}$, we define the Lagrangian function as follows:
\begin{eqnarray}
	&&	{\cal{L}} = \sum_{m=1}^M \lambda_m\left( \frac{\left(\frac{\mu}{a_m}\right)^{\frac{-1}{b_m}}}{\min\{\pi_{m,c}(\text{SINR}_m^\text{COM}),\pi_{s}(t_S) \}} - \tau_m\right) \nonumber \\
	&&	+ \lambda_{M+1}\left(\sum_{m=1}^M \tau_m - T \right) +\mu.
\end{eqnarray} 
The stationarity conditions are given by
\begin{subequations}
	\begin{eqnarray}
		&&  \frac{\partial{\cal{L}}}{\partial \mu} = \sum_{m=1}^M \lambda_m\left(\frac{-\left(\frac{\mu}{a_m}\right)^{\frac{-1}{b_m}-1}}{b_m\min\{\pi_{m,c}(\text{SINR}_m^\text{COM}),\pi_{s}(t_S) \}}\right) \nonumber \\
		&& +1=0,   \\ 
		&& \frac{\partial {\cal{L}}}{ \partial \tau_m} = -\lambda_m + \lambda_{M+1} = 0,\,\,\forall m. \label{K1}
	\end{eqnarray}
\end{subequations}  
The feasibility conditions of primal and dual problems are given by
\begin{subequations}
	\begin{eqnarray}
		&&\!\!\!\!\! \lambda_1,\lambda_2,\cdots,\lambda_{M+1} \ge 0, \\ && \!\!\!\!\! \frac{\left(\frac{\mu}{a_m}\right)^{\frac{-1}{b_m}}}{\min\{\pi_{m,c}(\text{SINR}_m^\text{COM}),\pi_{s}(t_S) \}} -\tau_m\le 0, \forall m,  \\
		&&\!\!\!\!\! \sum_{m=1}^M \tau_m -T \le 0,  
	\end{eqnarray}
\end{subequations}
The complementary slackness conditions are given by
\begin{subequations}
	\begin{eqnarray}
		&&\!\!\!\!\!\!\!\!\!\!\!\!\!\!\!\!\!\! \lambda_m\left( \frac{\left(\frac{\mu}{a_m}\right)^{\frac{-1}{b_m}}}{\min\{\pi_{m,c}(\text{SINR}_m^\text{COM}),\pi_{s}(t_S) \}} - \tau_m\right)=0, \forall m \label{K2} \\ 
		&& \!\!\!\!\!\!\!\!\!\!\!\!\!\!\!\!\!\! \lambda_{M+1}\left( \sum_{m=1}^M \tau_m -T  \right) = 0.
	\end{eqnarray}
\end{subequations}
For KKT solutions $\mu^*,\tau_m^*,\,\forall m$, we notice that 
\begin{equation}
	\sum_{m=1}^M \tau_m^* -T=0, \label{K3}
\end{equation}
and $\lambda_{M+1} > 0$. Otherwise, if  $\sum_{m=1}^M \tau_m^* -T < 0$ and $\lambda_{M+1} = 0$, we can keep deceasing the objective $\mu$ by enlarging $\tau_m^*,\,\forall m$. Due to $\lambda_{M+1} > 0$, we have $\lambda_m > 0,\,\forall m$, from \eqref{K1}. This implies that 
\begin{equation}
	\frac{1}{\min\{\pi_{m,c}(\text{SINR}_m^\text{COM}),\pi_{s}(t_S) \}}\left(\frac{\mu}{a_m}\right)^{\frac{-1}{b_m}} = \tau_m, \,\,\,\forall m, \label{K4}
\end{equation}
due to \eqref{K2}. Substituting \eqref{K4} into \eqref{K3}, we therefore have
\begin{equation}
	\sum_{m=1}^M \frac{1}{\min\{\pi_{m,c}(\text{SINR}_m^\text{COM}),\pi_{s}(t_S) \}}\left(\frac{\mu}{a_m}\right)^{\frac{-1}{b_m}} = T.
\end{equation}	

 \end{appendices}
\end{document}